\documentclass[aps,twocolumn,superscriptaddress]{revtex4-1}

\usepackage[final]{graphicx}
\usepackage{multirow}
\usepackage{xcolor}
\usepackage{units}
\usepackage{mathrsfs}
\usepackage{mathptmx, textcomp}
\usepackage{amssymb, amsmath}
\usepackage{ragged2e}
\usepackage{braket}
\usepackage[final,colorlinks,urlcolor=blue ,citecolor=blue,linkcolor=blue]{hyperref}

\usepackage[]{multibib} 
\newcites{methods}{The other list}

\usepackage{caption}
\DeclareCaptionLabelSeparator{vline}{$~\mathbf{\vert}$}
\makeatletter
\def\justified{
  \let\\\@normalcr%
  \@rightskip\z@skip \rightskip\@rightskip%
  \leftskip\z@skip%
  \parindent 0em\relax%
  \setlength{\parfillskip}{0pt plus 1fil}}%
\DeclareCaptionJustification{justified}{\justified}

\DeclareCaptionType{extfig}
\renewcommand{\figurename}{\hspace*{-.325em}Fig.}

\begin{document}

\title{Phase coherence in out-of-equilibrium supersolid states of ultracold dipolar atoms}

\author{P.~Ilzh\"ofer}
\affiliation{%
 Institut f\"{u}r Quantenoptik und Quanteninformation, \"Osterreichische Akademie der Wissenschaften, 6020 Innsbruck, Austria
}%
\author{M.~Sohmen}
\affiliation{%
 Institut f\"{u}r Quantenoptik und Quanteninformation, \"Osterreichische Akademie der Wissenschaften, 6020 Innsbruck, Austria
}%
\affiliation{%
 Institut f\"{u}r Experimentalphysik und Zentrum f\"{u}r Quantenoptik,\\ Universit\"{a}t Innsbruck, Technikerstra\ss e 25, 6020 Innsbruck, Austria
}%
\author{G.~Durastante}
\affiliation{%
 Institut f\"{u}r Quantenoptik und Quanteninformation, \"Osterreichische Akademie der Wissenschaften, 6020 Innsbruck, Austria
}%
\affiliation{%
 Institut f\"{u}r Experimentalphysik und Zentrum f\"{u}r Quantenoptik,\\ Universit\"{a}t Innsbruck, Technikerstra\ss e 25, 6020 Innsbruck, Austria
}%
\author{C.~Politi}
\affiliation{%
 Institut f\"{u}r Quantenoptik und Quanteninformation, \"Osterreichische Akademie der Wissenschaften, 6020 Innsbruck, Austria
}%
\author{A.~Trautmann}
\affiliation{%
 Institut f\"{u}r Quantenoptik und Quanteninformation, \"Osterreichische Akademie der Wissenschaften, 6020 Innsbruck, Austria
}%
\author{G.~Morpurgo}
\affiliation{%
 DQMP, University of Geneva, 24 Quai Ernest-Ansermet, CH-1211 Geneva, Switzerland
 }%
\author{T.~Giamarchi}
\affiliation{%
 DQMP, University of Geneva, 24 Quai Ernest-Ansermet, CH-1211 Geneva, Switzerland
 }%
\author{L.~Chomaz}
\affiliation{%
 Institut f\"{u}r Experimentalphysik und Zentrum f\"{u}r Quantenoptik,\\ Universit\"{a}t Innsbruck, Technikerstra\ss e 25, 6020 Innsbruck, Austria
}%
\author{M.~J.~Mark}
\affiliation{%
 Institut f\"{u}r Quantenoptik und Quanteninformation, \"Osterreichische Akademie der Wissenschaften, 6020 Innsbruck, Austria
}%
\affiliation{%
 Institut f\"{u}r Experimentalphysik und Zentrum f\"{u}r Quantenoptik,\\ Universit\"{a}t Innsbruck, Technikerstra\ss e 25, 6020 Innsbruck, Austria
}%
\author{F.~Ferlaino\hspace*{4mm}\textsuperscript{,*}\hspace*{-6mm}}\affiliation{%
 Institut f\"{u}r Quantenoptik und Quanteninformation, \"Osterreichische Akademie der Wissenschaften, 6020 Innsbruck, Austria
}%
\affiliation{%
 Institut f\"{u}r Experimentalphysik und Zentrum f\"{u}r Quantenoptik,\\ Universit\"{a}t Innsbruck, Technikerstra\ss e 25, 6020 Innsbruck, Austria
}%

\date{\today}

\maketitle

\noindent\textbf{A supersolid is a fascinating phase of matter, combining the global phase coherence of a superfluid with hallmarks of solids, e.\,g.~a spontaneous breaking of the translational symmetry. Recently, states with such counter-intuitive properties have been realized in experiments using ultracold quantum gases with strong dipolar interactions. Here, we investigate the response of a supersolid state to phase excitations which shatter the global phase coherence. After the creation of those excitations, we observe a rapid re-establishment of a global phase coherence, suggesting the presence of a superfluid flow across the whole sample and an efficient dissipation mechanism. We are able to identify a well-defined region where rephasing occurs, indicating the phase boundary between the solid-like and the supersolid phase. Our observations call for the development of theoretical descriptions able to capture the non-equilibrium dynamics in the recently discovered supersolid states of quantum matter.}

The notion of phase coherence lies at the foundation of quantum physics. It is considered a master property in understanding many-body quantum phenomena~\cite{CohenTannoudji:2011,Svistunov:2015}, ranging from superfluidity and the Josephson effect to the more applied examples of matter-wave interference, atom lasing processes, and quantum transport in mesoscopic and macroscopic systems. A coherent state can be described in terms of single amplitude and phase fields. However, the phase itself is not a physical observable and the study of coherence relies on measurements of phase differences between a set of coherent matter waves. In the context of atomic Bose--Einstein condensates (BECs), sets of spatially separated clouds have been created, for instance, by splitting a BEC into two or more parts or by loading an ultracold gas into an optical lattice or into a double-well potential~\cite{Bloch2008r}. 

The reverse process, i.\,e.~in-trap merging of BECs, and the related study of the phase evolution, is a much less explored and more subtle problem, invoking, for instance, the growth of thermal correlations in isolated systems~\cite{Langen:2013}, the complex interaction-mediated collapse and revival of many-body coherence~\cite{Wright96car}, dissipative dynamics~\cite{Jo:2007}, or even the exponential growth of unstable modes and topological defects in connection with the Kibble-Zurek mechanism~\cite{Scherer07vfb,Aidelsburger:2017,delCampo:2014}. Despite important theoretical and experimental progress, no generic framework exists yet to understand the quantum-phase evolution and relaxation dynamics in quantum many-body systems out of equilibrium~\cite{Langen:2015}.

The physical understanding of the dynamical re-establishment of coherence remains even more elusive for many-body quantum states that feature a spontaneous breaking of the translational symmetry. Prime examples are the supersolid states. For a long time mainly considered a theoretical notion~\cite{Andreev:1969,Chester:1970,Leggett:1970,Boninsegni:2012}, such states have been recently observed in quantum gases~\cite{Li:2017, Leonard:2017,Bottcher2019,Tanzi:2019,Chomaz:2019}. These supersolids can be seen as coherent matter waves with short-wavelength modulations -- shorter than the system size. Remarkably, in dipolar quantum gases the density modulation is not imprinted by external fields but truly emerges from the many-body interactions between atoms. Here, the symmetry breaking is driven by the interplay between short- (contact) and long-range (dipolar) interactions~\cite{Lu2015sds,Roccuzzo:2019,Youssef2019,Wenzel:2017,Baillie:2018,Cinti2017} and is connected to the softening of the roton mode in the excitation spectrum~\cite{Santos:2003,Chomaz:2018}. 

\begin{figure*}[t]
\centering
\includegraphics[width=\textwidth]{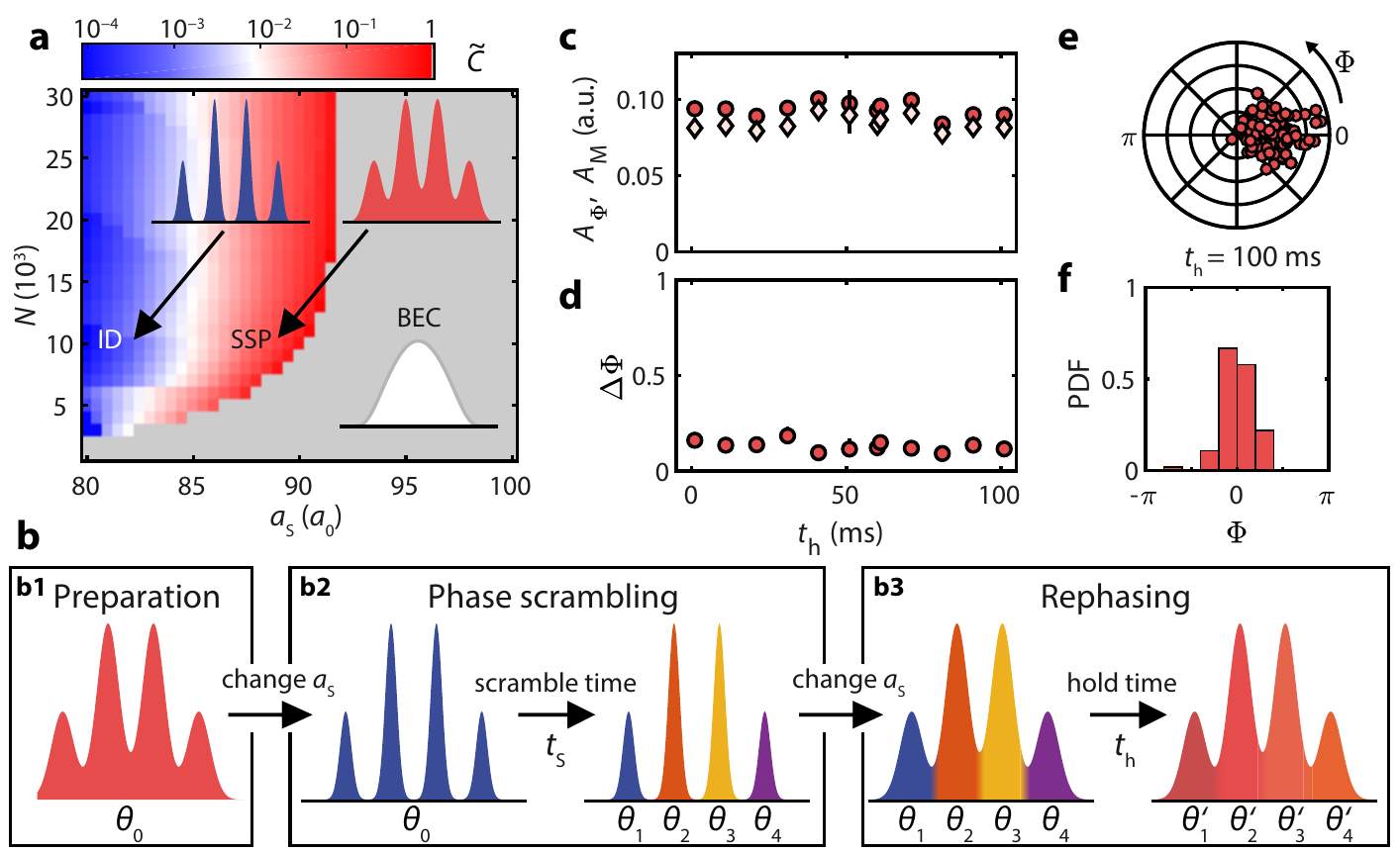} 
\caption{\label{fig1}
\textbf{Phase diagram, experimental sequence, and starting conditions.} \textbf{a,} Ground-state phase diagram for our cigar-shaped trapped $^{164}$Dy gas. The color map shows the values of $\widetilde{C}$. The grey color indicates a non-modulated BEC, while the red and blue regions correspond to a SSP and ID phase, respectively. The insets show illustrations of the density profiles along the weak axis for the different phases. \textbf{b,} Illustration of the phase scrambling sequence: Starting from a SSP (b1), we reduce $a_\text{s}$ to enter the ID regime (b2). During $t_\text{S}$, the phases of the droplets can evolve differently, leading to a phase scrambling between the individual droplets. Eventually, we jump $a_\text{s}$ back to its initial value, re-entering the supersolid regime (b3), where we study the time evolution of the global phase coherence. \textbf{c,} Amplitudes $A_\text{M}$ and $A_\Phi$, and \textbf{d,} $\Delta\Phi$ for our evaporatively cooled SSP plotted over $t_\text{h}$. Each data point is derived from $q=80$--$90$ individual experimental realizations. The error bars (almost covered by plot symbols) are the one-$\sigma$ confidence intervals calculated using a bias-corrected accelerated bootstrapping analysis (Methods)~\cite{Efron:1987}. \textbf{e,} Polar scatter plot for $P_i$ and \textbf{f,} histogram of the probability density function (PDF) for $\Phi_i$ at $t_\text{h}=100\,$ms.}
\end{figure*}

The density modulation is predicted to robustly survive both in the limit of infinite system size~\cite{Roccuzzo:2019} and trapped quantum gases~\cite{Bottcher2019,Chomaz:2019}. In the latter case, for a given trap geometry, the modulation contrast can be controlled by tuning the scattering length $a_\text{s}$~-- parametrizing the contact interaction --~or changing the atom number, $N$, in the system. The different quantum phases of a cigar-shaped dipolar quantum gas with vertical dipole orientation are shown in Fig.\,\ref{fig1}a. The phase diagram is constructed by numerically solving the extended Gross-Pitaevskii equation (eGPE; Methods)~\cite{Chomaz:2019}, describing our trapped system and including the recently discovered quantum-fluctuation-driven stabilization mechanism~\cite{Igor:2016,Chomaz:2016,Waechtler:2016,Bisset:2016}. The color map encodes the strength of the density modulation via the number $\widetilde{C}=1-C$, with ${C=(n_\text{max}-n_\text{min})/(n_\text{max}+n_\text{min})}$ the dimensionless modulation contrast. Here, $n_\text{max}$ ($n_\text{min}$) is the density maximum (minimum) in the central region of the calculated in-situ density distribution. $\widetilde{C}$~equals unity for a non-modulated and zero for a fully modulated state.  


The phase diagram shows three distinct regions; see Fig.\,\ref{fig1}a. For large enough $a_\text{s}$, the system is a non-modulated dipolar BEC with $\widetilde{C}=1$~(grey region). By lowering $a_\text{s}$, the influence of the dipolar interaction increases. When reaching a critical value of $a_\text{s}$, the system undergoes a phase transition to a supersolid phase~(SSP). Here, a density modulation with $\widetilde{C}<1$~appears in the ground-state density profile (red region). By further lowering $a_\text{s}$, the system evolves into an array of independent droplets~(ID) with an exponentially vanishing density link between individual droplets and $\widetilde{C}$ approaching zero (blue region). 


Recent experiments have shown a connection between the strength of the density modulation and the coherence properties of the system, revealing a clear difference between the SSP and ID phase~\cite{Bottcher2019,Tanzi:2019,Chomaz:2019}. In the SSP, a global phase is present along the whole system, whereas, in the ID case, phase coherence is absent. The latter behavior can be understood by considering that any fluctuation or excitation within a single isolated droplet will drive an independent evolution of the phases, which cannot lock to each other since particle flow is absent~\cite{Bloch2008r}. This type of dephasing has been studied in split BECs and atomic Josephson-junction arrays~\cite{Bloch2008r}.

While the phase evolution when moving from a SSP to an ID can be understood intuitively, highly fundamental and non-trivial questions arise when considering the opposite route, i.\,e.~when phase-incoherent isolated droplets are linked back together. 
First, will the out-of-equilibrium system spontaneously re-establish phase coherence? And, if yes, will it relax into its supersolid ground state or reach a quasi-stationary state? Second, which mechanism sets the rephasing timescale? Finally, whereto will the excitation energy be dissipated? Many-body quantum descriptions, as e.\,g.~a standard eGPE approach, are often inherently phase coherent and thus cannot capture such types of non-equilibrium dynamics. 


Here, we take first steps to experimentally answer those questions by studying the out-of-equilibrium phase dynamics of a supersolid state after a \emph{phase-scrambling} excitation. Our excitation scheme relies on an interaction quench and exploits the different coherence characters of the SSP and ID phase, as illustrated in Fig.\,\ref{fig1}b. 
In particular, after preparing a dipolar quantum gas in the SSP via direct evaporative cooling~(b1), we drive the system into the ID regime by lowering $a_\text{s}$~(b2). Here, we observe that the phase coherence gets quickly lost while the system remains density modulated.
When going back to the parameter regime where the SSP is again the ground state~(b3), we observe efficient rephasing dynamics, re-establishing the global phase coherence of the supersolid. Our measurements indicate the presence of superfluid flow with particles delocalizing across the density modulated gas as well as a dynamical mechanism dissipating the created phase excitations.


As starting point for the experiments, we produce the initial supersolid state by direct evaporative cooling from a thermal sample. As demonstrated in Ref.\,\cite{Chomaz:2019}, this is a powerful approach to create a long-lived supersolid state with a high degree of phase coherence. For the present work, our supersolid state contains about $N=1.4\times 10^4$ $^{164}$Dy atoms and is confined in an axially elongated optical-dipole trap of harmonic frequencies $\omega_{x,y,z}=2\pi\times\left(225,37,165\right)\,\text{s}^{-1}$. During the whole evaporation sequence, we apply a vertical magnetic field of $B=2.430(4)\,$G to set the dipole orientation and the desired $a_\text{s}$-value in the SSP region.

Our investigation relies on the ability to probe the system's phase coherence and density modulation, whose co-existence is a hallmark of supersolidity. To this aim, we developed an analysis based on matter-wave-interference~\cite{Chomaz:2019,Hadzibabic,takeda1982ftm,Kohstall2011,Chomaz:2015eoc,Bottcher2019}, which is capable of capturing the degree of phase coherence and the density-modulation strength (Methods). In brief, for each individual experimental realization $i$, we take an absorption image after a time-of-flight (TOF) expansion, which exhibits an interference pattern in case of an in-situ density modulation. Via Fourier transform, we extract the phasor $P_i=\rho_i\cdot \text{e}^{-\text{i}\cdot\Phi_i}$, revealing the amplitude $\rho_i$ and phase $\Phi_i$ at the spatial frequency of the interference pattern. Whereas a single $P_i$ characterises the degree of density modulation, the statistical average over an ensemble $q$ of many realizations reveals information about the global phase coherence. We calculate the phase amplitude, $A_\Phi=|\langle P_i\rangle|$, and the density-modulation amplitude, $A_\text{M}=\langle |P_i|\rangle$, as well as the circular phase variance $\Delta\Phi=1-\frac{1}{q} \sqrt{\left(\sum_{i=1}^{q}\cos\left(\Phi_i\right)\right)^2+\left(\sum_{i=1}^{q} \sin\left(\Phi_i\right)\right)^2}$~\cite{Fisher1995}. We note that for a perfect supersolid (resp.~ID) state and in the limit $q\to\infty$, $A_\Phi=A_\text{M}>0$ (resp.~$A_\Phi=0$, $A_\text{M}>0$) and $\Delta\Phi=0$ (resp.~$1$).

\begin{figure}[t]
\centering
\includegraphics[width=\columnwidth]{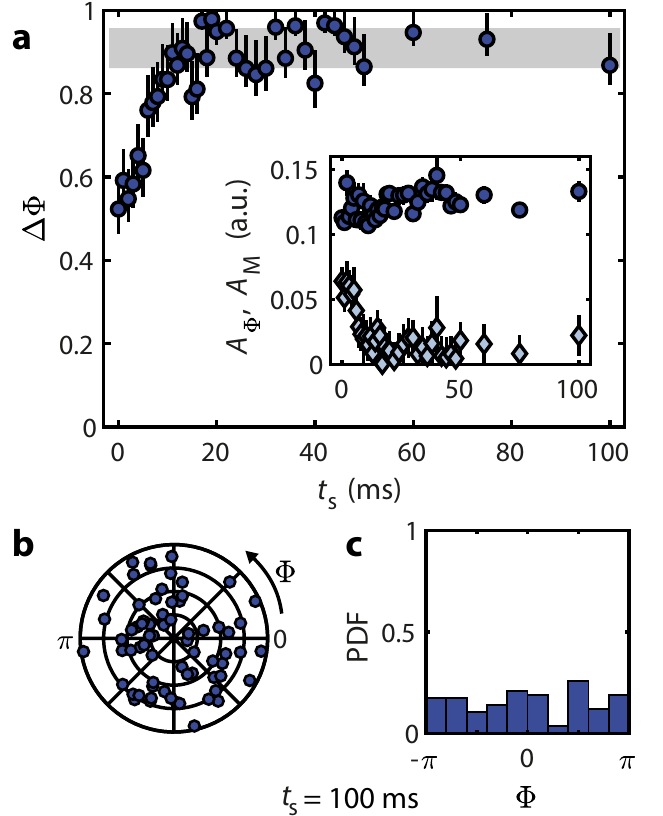} 
\caption{\label{fig2}
\textbf{Phase scrambling.} \textbf{a,} $\Delta\Phi$ as a function of $t_\text{S}$ for the ID phase at $1.65\,$G ($a_\text{s,ID}=77\,a_0$). Each point is derived from $q=90$--$100$ independent experimental realizations. The error bars are the one-$\sigma$ confidence intervals calculated using a bias-corrected accelerated bootstrapping analysis (Methods)~\cite{Efron:1987}. The grey shaded area indicates the theoretical one-$\sigma$ confidence interval for $\Delta\Phi$ using the same sample size and a uniformly random phase. The inset shows the according $A_\Phi$ (cyan) as well as $A_\text{M}$ (blue). \textbf{b,} Polar scatter plot for $P_i$ and \textbf{c,} histogram of the PDF for $\Phi_i$ at $t_\text{S}=100\,$ms.}
\end{figure}


To demonstrate the power of this analysis, we apply it to our initial state, whose supersolid properties have been previously investigated~\cite{Chomaz:2019}. As shown in Fig.\,\ref{fig1}c--d, $A_\Phi$, $A_\text{M}$ and $\Delta\Phi$ are roughly constant during holding times $t_\text{h}$ up to $100\,$ms. We observe almost equal values for $A_\Phi$ and $A_\text{M}$ and a mean value $\langle\Delta\Phi\rangle=0.142(8)$, confirming a high degree of global phase coherence for our density-modulated initial state. 
For $t_\text{h}=100\,$ms, we also show a polar plot of $P_i$ (Fig.\,\ref{fig1}e) as well as the corresponding histogram for $\Phi_i$ (Fig.\,\ref{fig1}f), both displaying a narrow distribution.

After preparing the initial supersolid state, we apply our phase-scrambling protocol; see Fig.\,\ref{fig1}b2. We ramp the $B$-field within $20\,$ms from $2.43\,$G (SSP) to $1.65\,$G (ID phase) and let the system evolve for a variable time $t_\text{S}$. Exploiting the magnetic-field tunability of $a_\text{s}$ via Feshbach resonances, the $B$-field ramp corresponds to a change from about $a_\text{s,SSP}=88\,a_0$ to $a_\text{s,ID}=77\,a_0$ (Methods). As shown in Fig.\,\ref{fig2}a, we observe a rapid initial increase of $\Delta\Phi$ on a time scale of $t_\text{S} \simeq 20\,$ms \footnote{Note that $\Delta\Phi \left(t_\text{S}=0\right)\simeq 0.5$. We attribute this large value to dephasing dynamics taking place already within the $B$-field ramp duration and the initial time of the TOF.}, after which $\Delta\Phi$ saturates close at a mean value of $\langle\Delta\Phi\rangle_{t_\text{S}\geq 30\,\text{ms}}=0.92(2)$. We note that the saturation value is not expected to reach unity because of our finite sample size ($q \simeq 95$). Indeed, it is comparable to the one calculated from a toy model, which considers a sample with the same $q$ and fully random (i.\,e.~uniformly distributed) phases (Methods).

Simultaneous to the increase of $\Delta\Phi$, we observe that $A_\Phi$ decreases quickly towards zero, while $A_\text{M}$ slightly increases. This behavior shows that the density modulation is maintained while losing global phase coherence; see inset. As expected, in the ID phase, quantum and thermal fluctuations as well as atom losses can give rise to a different time evolution for the phases of the individual droplets. Apparently the vanishing small density overlap between droplets ($\widetilde{C}\simeq 0$) prevents an efficient phase locking, resulting in the observed loss of global phase coherence. 


\begin{figure}[t]
\centering
\includegraphics[width=\columnwidth]{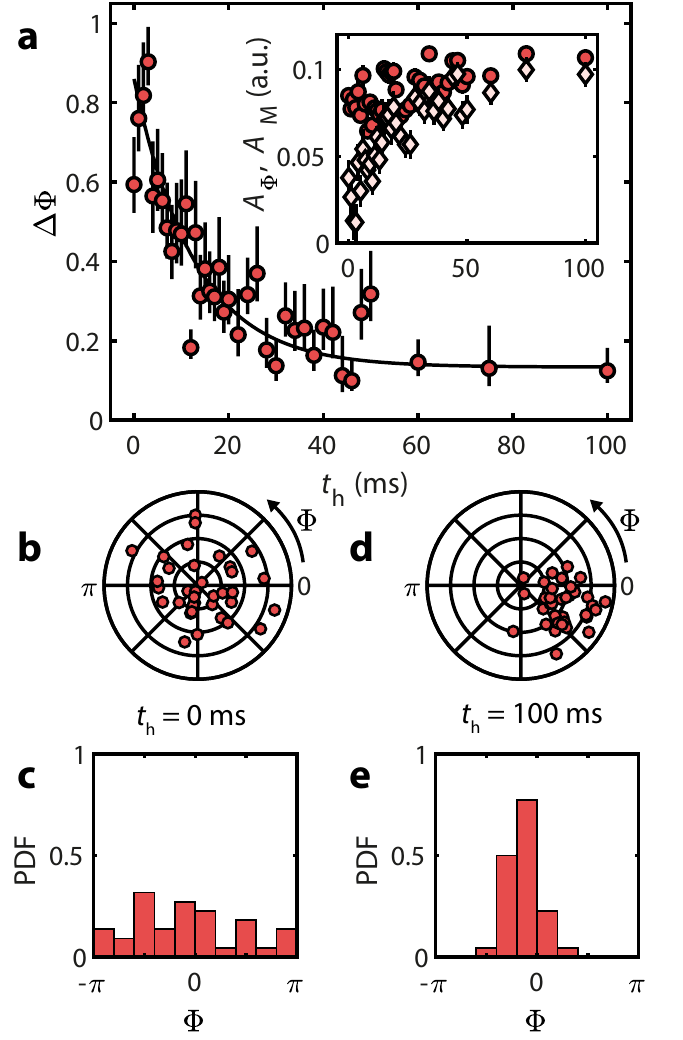} 
\caption{\label{fig3}
{\bf Rephasing dynamics.} \textbf{a,} $\Delta\Phi$ as a function of $t_\text{h}$ after a jump from the ID phase back to the SSP regime at $2.43\,$G ($a_\text{s,SSP}=88\,a_0$). For each point $q=66$--$74$. The error bars are the one-$\sigma$ confidence intervals calculated using a bias-corrected accelerated bootstrapping analysis (Methods)~\cite{Efron:1987}. The solid black line is an exponential fit to guide the eye. 
The inset shows the according $A_\Phi$ (light red) as well as $A_\text{M}$ (red). \textbf{b,} Polar scatter plot for $P_i$ and \textbf{c,} histogram of the PDF for $\Phi_i$ at $t_\text{h}=0\,$ms and at $t_\text{h}=100\,$ms (\textbf{d}--\textbf{e}).}
\end{figure}


We now move to the core of our experiment and investigate the phase-relocking after the phase scrambling. We set $a_\text{s}$ back to its initial value, i.\,e.~where the system's ground state is a supersolid, by a $B$-field jump, and study the system's evolution; see Fig.\,\ref{fig1}b3. As shown in Fig.\,\ref{fig3}a, we observe first a rapid reduction of $\Delta\Phi$, occurring in the first $20\,$ms, and then a much slower dynamics with $\Delta\Phi$ saturating to $\langle\Delta\Phi\rangle_{t_\text{h}\geq 30\,\text{ms}}=0.20(2)$. Accordingly, $A_\Phi$ approaches $A_\text{M}$ on the same time scale, whereas $A_\text{M}$ remains nearly constant. This re-establishment of global phase coherence is further illustrated with individual polar scatter plots and histograms in Fig.\,~\ref{fig3}, confirming a reduction of the phase distribution's width with increasing $t_\text{h}$. 

Our system of multiple superfluid parts with different phases interconnected via weak links is reminiscent of a Josephson-Junction array~(JJA)~\cite{fazio:2001}, opening the question whether a JJA framework can capture the main ingredients of our system's dynamics. Although our array of droplets is soft, meaning that the droplets' shape and their distance change with $a_\text{s}$, we construct a simple model in terms of a one-dimensional array of coupled grains (Methods). This is justified as the strongest effect of the change of the system's state with $a_\text{s}$ is the change of the wavefunction overlap between the droplets, i.\,e.~the tunneling rate. Using this model, we simulate quenches of the tunneling rate and look at the time evolution of the correlation function of the phases in the array, which corresponds to the experimental observable $A_\Phi/A_\text{M}$. 

The model gives dephasing and rephasing dynamics, similar to the observations of Fig.\,\ref{fig2}-\ref{fig3}. A more quantitative description goes beyond the scope of this paper. It would require (i) to find proper relations between the parameters of the JJA model and the real system, (ii) to achieve a macroscopic modelling of the dissipation mechanisms by including coupling with a thermal bath and/or with the excited droplet modes~\cite{Torre:2013}, or even (iii) to go beyond the hard-grain model. Even in experiments with non-dipolar coupled quasi-condensates, realizing a case closer to an ideal JJA, the observed phase dynamics and full phase-locking have no theoretical explanation up to now~\cite{Pigneur:2018}. Another important ingredient in the phase relaxation dynamics is the phase defects formed at the boundaries between the distinct grains when they merge~\cite{Jo:2007,Frantzeskakis:2010,Aidelsburger:2017,delCampo:2014}. These defects, forming e.\,g.~solitons, are expected to propagate and interact with each other and with excitations from the thermal bath, and thus eventually decay.

\begin{figure*}[t]
\centering
\includegraphics[width=\textwidth]{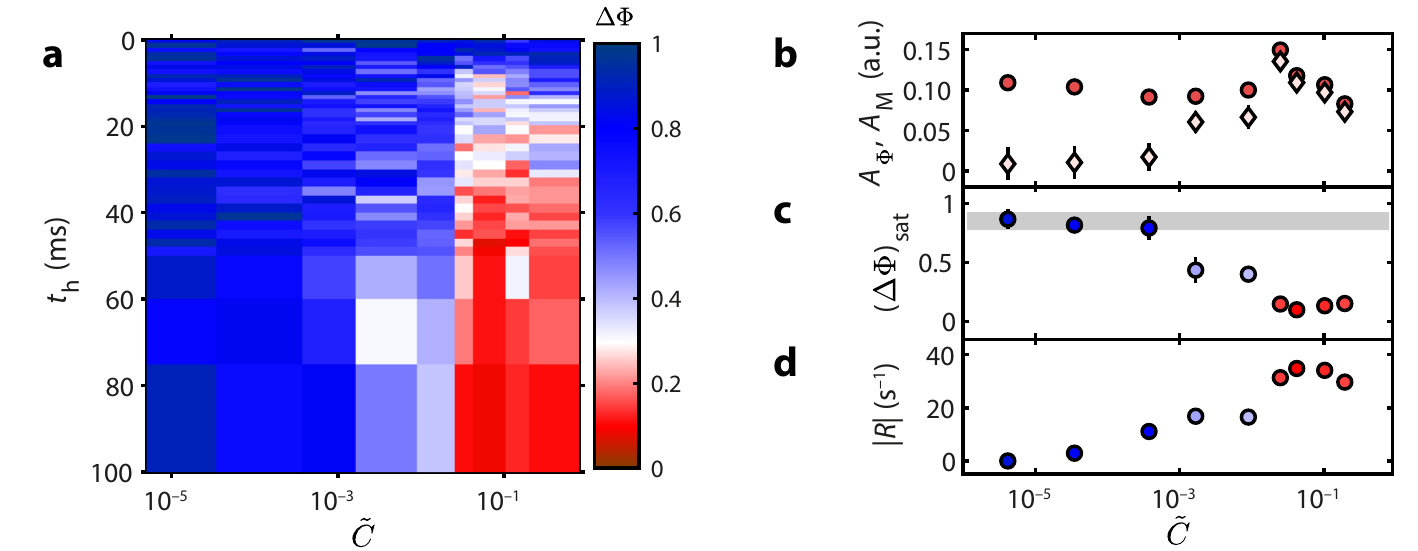} 
\caption{\label{fig4}
{\bf Time traces of the phase dynamics and their characterization.} \textbf{a,} Temporal evolution of $\Delta\Phi$ (color map) after the complete phase scrambling sequence plotted as a function of $t_\text{h}$ and $\widetilde{C}$. For each $t_\text{h}$ we record $q\geq 35$ individual experimental realizations. In the red region ($\Delta\Phi\simeq 0$) the system has recovered its global phase coherence, while for the blue one ($\Delta\Phi\simeq 1$) no global phase coherence is present. \textbf{b,} $A_\Phi$ (light red) and $A_\text{M}$ (red), \textbf{c,} saturation value $\left(\Delta\Phi\right)_\text{sat}$ and \textbf{d,} rephasing rate $|R|$ as a function of $\widetilde{C}$. $A_\Phi$,  $A_\text{M}$, and $\left(\Delta\Phi\right)_\text{sat}$ are the mean values at long $t_\text{h}$. The grey shaded area indicates the theoretical one-$\sigma$ confidence interval for $\Delta\Phi$ using the same sample size as the experiment and a uniformly random phase.}
\end{figure*}


To further investigate the role of the density links among droplets -- i.\,e.~the Josephson coupling --, we study the rephasing dynamics as a function of the theoretically calculated $\widetilde{C}$; see Fig.\,\ref{fig1}a. Although our system is out of equilibrium, we use the ground-state quantity $\widetilde{C}$ as an estimate for the strength of the density link~\cite{Bottcher2019, Tanzi:2019,Chomaz:2019}. For each value of $a_\text{s}$, we assign $\widetilde{C}$ and record the time trace of $\Delta\Phi$ for different $t_\text{h}$. As shown in Fig.\,\ref{fig4}a, we see different rephasing dynamics depending on $\widetilde{C}$. In the case of small $\widetilde{C}$, associated with the ID regime, no rephasing occurs with $\Delta\Phi$ remaining large ($>0.5$) for all $t_\text{h}$ (blue region). As $\widetilde{C}$ and thus the link strength increases, the system starts to rephase with $\Delta\Phi$ approaching a small saturation value ($\approx 0.15$) at long evolution times (red region). 

The time traces clearly show the existence of two regimes for $\Delta\Phi$, one in which phase re-locking occurs and one in which the system remains incoherent. To further investigate these regimes and their interface, we study the long-time dynamics of $A_\Phi$ and $A_\text{M}$ and record their asymptotic values. 
As shown in Fig.\,\ref{fig4}b, $A_\text{M}$ remains large and shows only slight variations over the full investigated range of $\widetilde{C}$. This indicates the persistence of density modulation in the system. In contrast, a striking change is found in the evolution of the ratio between $A_\Phi$ and $A_\text{M}$. At large $\widetilde{C} >0.01$, $A_\Phi$ and $A_\text{M}$ are nearly equal. This shows the re-establishement of a global phase coherence, and the relaxation towards a SSP. Differently, at small $\widetilde{C}<0.001$, $A_\Phi$ nearly vanishes while $A_\text{M}$ remains large and almost constant, evidencing a final phase-incoherent state (ID regime). At intermediate $\widetilde{C}$, $A_\Phi$ and $A_\text{M}$ show an in-between behavior with $A_\Phi$ smaller than $A_\text{M}$ but non-vanishing, showing a partial recoherence of the state.
These three distinct behaviors are also reflected in the asymptotic values of $\Delta\Phi$ (Fig.\,\ref{fig4}c), showing full recoherence ($\Delta\Phi\approx 0.15$) in the supersolid regime, persistence of a full incoherence ($\Delta\Phi\approx 0.9$) in the ID regime, and partial recoherence ($\Delta\Phi\approx 0.5$) in the intermediate regime. 

A further question is whether the value of $\widetilde{C}$, i.\,e.~the different regimes, also dictates the speed of the rephasing dynamics. To explore this aspect, we study the early time dynamics of $\Delta\Phi$ by performing a linear fit to the data for $t_\text{h}\leq 20\,$ms. The extracted slope characterizes the initial rephasing rate $|R|$. As shown in Fig.\,\ref{fig4}d, in the supersolid regime, we always record a large rephasing rate, which remarkably is roughly independent of $\widetilde{C}$, with $|R|\approx 30\text{s}^{-1}$. This value is comparable to the weak-axis trap frequency, $\omega_y/2\pi$, and compatible with the time needed for a sound wave or soliton to propagate along the system~\cite{Natale:2019,Aidelsburger:2017}. In contrast, when crossing from the supersolid to the intermediate regime, we observe a sudden decrease of $|R|$ by almost a factor of two. Evolving from this intermediate regime to the ID, $|R|$ continuously decreases with decreasing $\widetilde{C}$ 
until it vanishes. While such a decrease of $|R|$ is consistent with a JJA picture in which the tunneling between the droplets dictates the rephasing dynamics, the underlying reason for a constant rephasing rate in the supersolid regime remains an open question. It might indicate the action of other mechanisms, related for instance to the soft nature of our JJA, or the formation and slow decay of phase defects in the array. 


In conclusion, we have reported the first study of the out-of-equilibrium dynamics of a dipolar supersolid after an interaction-driven phase excitation that fully randomizes the phases. In the SSP regime, we have demonstrated that the system re-establishes a high-degree of phase coherence on the timescale of one trap period by almost perfect rephasing. When tunneling is suppressed by a too weak density link across our spontaneously-modulated quantum state, the rephasing substantially slows down at a rate depending on the tunneling and eventually ceases in the deep ID regime. Our observations might shed new light on the properties of the particle flow in a SSP and its superfluid properties, whose general understanding is still elusive. Future experimental works, combined with advanced out-of-equilibrium theoretical models, will be crucial to understand the relaxation dynamics and dissipation mechanisms in isolated and open supersolid states of quantum matter.


\bibliographystyle{naturemag}
\bibliography{references}


\begin{itemize}

\item We are grateful to S.~Erne, J.~Schmiedmayer and the ERBIUM team for insightful discussions. We acknowledge R.~M.~W.~van Bijnen for developing the code for our eGPE ground-state simulations. This work is financially supported through an ERC Consolidator Grant (RARE, No.\,681432), a NFRI Grant (MIRARE, No.\,\"OAW0600) from the Austrian Academy of Science and a DFG/FWF (FOR 2247/PI2790), and by the Swiss National Science Fundation under Division II. M.~S.~and G.~D.~acknowledge support by the Austrian Science Fund FWF within the DK-ALM (No.\,W1259-N27). A.~T.~acknowledges support by the Austrian Science Fund FWF within the Lise-Meitner-Programm (No.\,M2683-N36). We also acknowledge the Innsbruck Laser Core Facility, financed by the Austrian Federal Ministry of Science, Research and Economy. Part of the computational results presented have been achieved using the HPC infrastructure LEO of the University of Innsbruck.

\item Competing Interests:
The authors declare no competing interests.
 
\item Author information: Correspondence to F.~F.~(e-mail: Francesca.Ferlaino@uibk.ac.at).

\item Author contributions: P.~I.~, G.~D.~and A~T.~conducted the experiment and collected the experimental data. M.~S.~and C.~P.~analyzed the data. M.~J.~M.~performed the eGPE ground-state simulations. G.~M.~and T.~G.~contributed to the theoretical analysis. All authors contributed to the writing of the paper. F.~F.~supervised the project.

\end{itemize}


\clearpage

\section{Methods}

\subsection{Phase diagram and contrast.}
Our numerical calculations of the ground-state phase diagram of a cigar-shaped $^{164}$Dy dipolar quantum gas follow the procedure described in our earlier works~\cite{Chomaz:2018,Chomaz:2019}. In brief, the calculations are based on minimizing the energy functional of the extended Gross-Pitaevskii equation (eGPE) using the conjugate-gradients technique~\citemethods{Ronen:2006}. The eGPE includes our anisotropic trapping potential, the short-range contact and long-range dipolar interactions at a mean-field level, as well as the first-order beyond-mean-field correction in the form of a Lee-Huang-Yang (LHY) term~\cite{Chomaz:2016,Waechtler:2016} and~\citemethods{Waechtler:2016b}. From the derived three-dimensional wavefunction $\psi(\mathbf{r})$ we calculate the one-dimensional in-situ density profile ${n(y)=\int |\psi(\mathbf{r})|^2\mathrm{d}x\mathrm{d}z}$. We evaluate the in-situ density contrast ${C=(n_\text{max}-n_\text{min})/(n_\text{max}+n_\text{min})}$ for profiles which feature density modulations by searching central extrema of $n(y)$ and determining the overall maximum~($n_\text{max}$) and minimum~($n_\text{min}$) value. For profiles without density modulation (ordinary BEC), we set $C = 0$. We use the quantity ${\widetilde{C}=1-C}$ to estimate the density link between the droplets, which is connected to the tunneling strength~\cite{Chomaz:2019}.

\subsection{Experimental sequence.}
We apply our phase scrambling protocol to the evaporatively cooled SSP of $^{164}$Dy atoms~\cite{Chomaz:2019}. For this, we initially load our 5-beam open-top magneto-optical trap (MOT) for $3\,$s and apply a MOT compression phase, which lasts $400\,$ms~\citemethods{Ilzhoefer2018}. We then load about $8\times 10^{6}$ atoms into a single-beam horizontal optical dipole trap (hODT), propagating along our $y$-axis. The hODT is derived from a $1064\,$nm focused laser beam. After loading, we apply forced evaporative cooling by exponentially reducing the optical power in the hODT for $0.9\,$s. Subsequently, we switch on a second ODT beam along the vertical $z$-axis to form a crossed ODT and continue with the last stage of evaporative cooling for $2\,$s~\citemethods{Trautmann2018}, until the SSP is reached. During the entire evaporation sequence, a magnetic field of $B = 2.43\,$G, pointing opposite to gravity along our $z$-direction, is maintained. The final trap geometry is cigar-shaped with harmonic frequencies $\omega_{x,y,z} = 2\pi\times(225,37,165)\,\text{s}^{-1}$. After the initial-state preparation (SSP), we apply our phase-scrambling protocol. For that, without any additional waiting time after the evaporative cooling, we change the $B$-field to $1.65\,$G deep in the ID regime. Here, we allow the system's global phase to freely evolve for $t_\text{S}=20\,$ms. We have explored two types of protocols: jumping, which results in an effective $\approx 1\,$ms change of the $B$-field due to the finite time response of the system, and ramping within $20\,$ms. We observe a similar scrambling behavior in $\Delta \Phi$ for both the jump and the ramp protocol. We complete our phase-scrambling sequence by jumping the $B$-field back to its initial value and by letting the system evolve for a variable hold time $t_\text{h}$.
Finally, we perform a matter-wave interference-type experiment during time-of-flight (TOF) expansion and record the resulting interference pattern by absorption imaging. A TOF duration of $26.5\,$ms ensures a mapping onto momentum space. The imaging beam propagates along $\tilde{x}$ in the horizontal $x$--$y$-plane at an angle of $\sim 45\,^\circ$ with respect to the weak trap axis along $y$. 


\subsection{Tuning the scattering length.}
To connect the experimental $B$-field values with the contact scattering length $a_\text{s}$, we use the well established formula for overlapping Feshbach resonances $a_\text{s}\left(B\right)=a_\text{bg}\prod_i \left(1-\Delta B_i/\left(B - B_{0,i}\right)\right)$~\citemethods{Chin2010fri}, with $B_{0,i}$ the poles, $\Delta B_i$ the corresponding distance from the pole to the zero-crossing and $a_\text{bg}$ the (local) background scattering length. We determine the poles and zero-crossings in our $B$-field region of interest by performing loss spectroscopy and thermalization measurements. Starting from a thermal cloud prepared at $2.55\,$G we first ramp the magnetic field to the final value within $5\,$ms, then lower the trap depth to its final value within $50\,$ms, and wait an additional hold time of about $400\,$ms. In absence of Feshbach resonances, we typically end up with a thermal gas of $5\times 10^5$ atoms with a temperature of about $500\,$nK. When scanning the magnetic field in our region of interest, we observe several atom loss features together with peaks in the atom cloud temperature, which we fit with gaussian functions to extract the positions of the poles $B_{0,i}$ and the widths $\Delta B_i$. 

The value of the background scattering length of $^{164}$Dy is a more subtle topic, as several measurements give varying values in the range between $60$--$100\,a_0$~\citemethods{FerrierBarbut:2018}. These measurements were using different methods (e.\,g.~cross-thermalization, theory-experiment comparisons of oscillation frequencies), different initial states (thermal gases and quantum droplets) and were performed at different magnetic fields. Especially the existence of very broad resonances at higher magnetic fields~\citemethods{Maier:2015} will affect the measured local background scattering lengths. Therefore, we set the value of $a_\text{bg}$ in such a way that the $B$-to-$a_\text{s}$ conversion reproduces the calculated critical scattering length $a_\text{s} = 91\,a_0$ at the experimentally estimated phase transition point between BEC and SSP around $2.5\,$G. This gives a value of $a_\text{bg}=73\,a_0$ which lies within the error bars of the latest published value of $a_\text{bg}=69(4)\,a_0$~\citemethods{FerrierBarbut:2018}. Extended Data Figure~\ref{fig:asVsB} shows the resulting calculated $B$-to-$a_\text{s}$ conversion from which we estimate $a_\text{s,SSP}=88\,a_0$ at $2.43\,$G in the SSP and $a_\text{s,ID}=77\,a_0$ at $1.65\,$G in the ID as used in the experiment.



\begin{extfig}[t]
\includegraphics[width=\columnwidth]{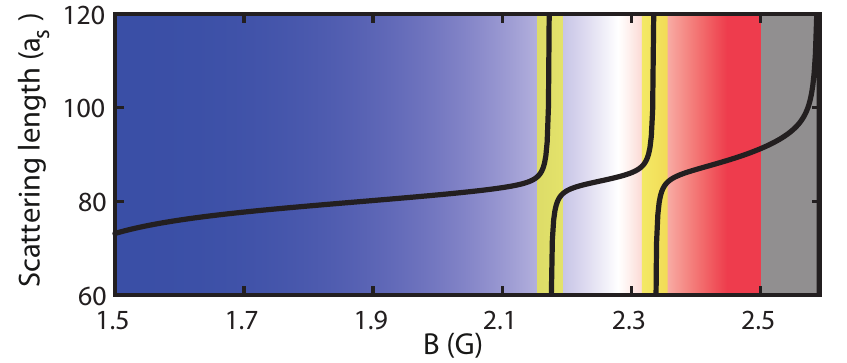}
\caption{\label{fig:asVsB}
\textbf{Estimated scattering length.} Calculated $B$-to-$a_\text{s}$ conversion for $^{164}$Dy. Red and blue shaded areas indicate the SSP and the ID region, respectively. The grey area indicates the BEC region, while the yellow areas indicate regions around the two narrow Feshbach resonances located at $2.174\,$G and $2.336\,$G where we observe increased atom loss. We estimate $a_\text{s,SSP}=88\,a_0$ in the SSP at $2.43\,$G and $a_\text{s,ID}=77\,a_0$ in the ID at $1.65\,$G.}
\end{extfig}

\subsection{Interference pattern analysis.}
Our analysis is similar to the one described in Ref.\,\cite{Chomaz:2019}. We record $q=30$--$100$ experimental repetitions for each parameter set $\mathcal{P}$.
Each recorded picture $i$ ($i=1\ldots q$) is processed by first subtracting the thermal background via a symmetric 2D-Gaussian fit to the wings of the density distribution. Next, we recenter the image of the degenerate cloud and integrate its central region, where the matter-wave interference signal is concentrated, along the $z$-direction within $\pm 2\,{\mu\text{m}}^{-1}$. We obtain a momentum density profile which we normalize by its sum. From such a momentum profile, a fast Fourier transformation (FFT) yields the 1D density profile $n_i\left(\tilde{y}\right)$. An in-situ density modulation in an atomic cloud will lead to side peaks in $n_i\left(\tilde{y}\right)$, symmetrically centred around the zero-momentum peak. 
To isolate the centre of this specific modulation, we calculate the incoherent and coherent means of $n_i\left(\tilde{y}\right)$, which we denote $n_\text{M}\left(\tilde{y}\right)= \langle|n_i\left(\tilde{y}\right)|\rangle_\mathcal{P}$ and $n_\Phi\left(\tilde{y}\right)= |\langle n_i\left(\tilde{y}\right)\rangle_\mathcal{P}|$, respectively.
The incoherent mean $n_\text{M}$ reflects the mean modulation amplitude of the cloud at the respective wavelength $\tilde{y}$. The coherent mean $n_\Phi\lesssim n_\text{M}$ if the phases of the interference pattern among the $q$ repetitions at the respective $\tilde{y}$ are roughly constant, and $n_\Phi \rightarrow 0$ (and hence $n_\Phi \ll n_\text M$) if the phases are random. Therefore, the most pronounced difference $n_\text{M}-n_\Phi$ is observed for the ID regime (see Extended Data Fig.\,\ref{fig:Sup_nMn-Phi}a).
From the maximum of this difference we read off the modulation wavelength (or `droplet distance') $\tilde{y} \equiv d$. 
The FFT phasors at $d$ we call $P_i=n_i\left(d\right)=\rho_i\cdot \text{e}^{-\text{i}\cdot\Phi_i}$, yielding sets $\{P_1,\ldots,P_q\}_\mathcal{P}$. 
To characterise the distribution of phases $\Phi_i$ within our sets, we calculate the circular variance $\Delta\Phi=1-\frac{1}{q} \sqrt{\left(\sum_{i=1}^{q}\cos\left(\Phi_i\right)\right)^2+\left(\sum_{i=1}^{q} \sin\left(\Phi_i\right)\right)^2}$~\cite{Fisher1995}. For a phase-coherent sample, and hence interference fringes stable within the envelope, $\Delta\Phi$ is small, whereas for an incoherent sample it approaches unity. 
To estimate the confidence intervals of our circular variance data we apply a bias-corrected accelerated bootstrapping scheme~\cite{Efron:1987} for each $\mathcal{P}$, resampling $10^6$ times from the respective $q$ experimental values.

\begin{extfig}[t]
\includegraphics[width=\columnwidth]{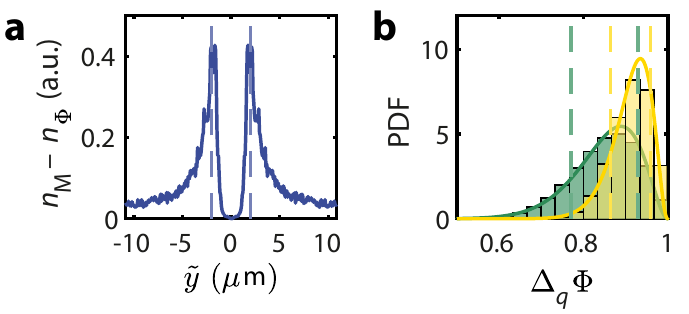}
\caption{\label{fig:Sup_nMn-Phi}
\textbf{Wavelength of the modulation and finite-sampling effect.} \textbf{a,} Difference between incoherent and coherent mean of the density profiles in the ID regime (1.65 G), peaking at the modulation wavelength $d \simeq \pm 2\,\mu$m (dashed lines). \textbf{b,} Histograms of $10^6$ realisations (each) for calculations of $\Delta_q\Phi$ from uniformly random phases $\Phi_i$, for $q = 35$ (green) and $q=100$ (yellow) draws, respectively. The dashed vertical lines reflect the confidence interval enclosing 68.3\,\% (`one $\sigma$') of the calculated values. The solid lines reflect a Beta distribution with same mean and variance as the drawn distributions of $\Delta_q\Phi$ (no free fit parameters).} 
\end{extfig}

\subsection{Effect of finite sample size}
Even the circular variance $\Delta\Phi$ of a sample of $q$ angles $\Phi_1,\ldots,\Phi_q$ drawn from a completely random distribution will approach unity only in the limit $q \rightarrow \infty$. To estimate the fully-incoherent limit of $\Delta\Phi$ for our finite $q$, we calculate $10^6$ values for $\Delta_q\Phi$, each for $q$ independent draws from a theoretical, uniform distribution in $[0, 2\pi)$. The histograms of $\Delta_q\Phi$ are shown in Extended Data Fig.\,\ref{fig:Sup_nMn-Phi}b. The indicated one-$\sigma$ confidence intervals are $[0.77, 0.93]$ for $q = 35$ and $[0.86, 0.96]$ for $q = 100$ draws.
We note that the histograms of $\Delta_q\Phi$ follow a Beta distribution~\citemethods{Evans:2000}, even if one generalizes the underlying distribution of phases $\Phi_i$ to a von Mises distribution, of which the uniform distribution is just a degenerate case.

\subsection{Interference pattern analysis and simple model of a droplet array}

For simplicity let us consider here that the state is made of $N_\text{D}$ identical droplets. In that case the total wavefunction of the system would be 
\begin{equation}
 \psi\left(x,y,z\right) = \sum_{j=1}^{N_\text{D}} f\left(x,y-R_j,z\right) \text{e}^{\text{i}\theta_j} 
\end{equation} 
where $R_j$ is the spatial coordinate of the $j$th droplet, $\theta_j$ is its phase, taken to be uniform over the droplet, and $f$ is the wavefunction of a single droplet localized around $y=0$. 
With such a wavefunction, the phasor extracted from one realization would be
\begin{equation} \label{eq:phasor0}
P_i=\int \mathrm{d}k_y \sum_{j_1,j_2=1}^{N_\text{D}} \text{e}^{ik_y\left(R_{j_1}-R_{j_2}-d\right)} \text{e}^{\text{i}\left(\theta_{j_1}-\theta_{j_2}\right)} |\tilde{f}\left(k_y\right)|^2
\end{equation} 
where $\tilde{f}$ is the Fourier transform of the function $f$ and $d$ is the distance between neighboring droplets $d=\langle R_{j+1}-R_{j}\rangle$. 
It simplifies in 
\begin{align} 
P_i&=\sum_{j_1,j_2=1}^{N_\text{D}} g\left(R_{j_1}-R_{j_2}-d\right)  \text{e}^{\text{i}\left(\theta_{j_1}-\theta_{j_2}\right)}\\
\label{eq:phasor}
&\approx g\left(0\right) \sum_{j}\text{e}^{\text{i}\left(\theta_{j+1}-\theta_{j}\right)}  
\end{align} 
with $g\left(y\right)$ the Fourier transform of $|\tilde{f}\left(k_y\right)|^2$, which is thus a peak function with a width of the order of the droplet size.

Formula~(\ref{eq:phasor}) yields 
\begin{equation}
A_\text{M}=\langle|P_i|\rangle_\mathcal{P} \approx \left(N_\text{D}-1\right) g\left(0\right),
\end{equation}
which is essentially independent on the phase relation between the droplets and shows only a weak dependence on the droplets' shape. Here $\langle.\rangle_\mathcal{P}$ denote the average over an ensemble of realizations $\mathcal{P}$. 

On the contrary the function $A_\Phi=|\langle P_i\rangle_\mathcal{P}|$ contains the average of the phases with 
\begin{equation} 
\frac{A_{\Phi}}{A_{\text{M}}}=\frac{|\langle P_i \rangle_\mathcal{P}|}{\langle|P_i|\rangle_\mathcal{P}} \simeq |\langle \langle \text{e}^{\text{i}\left(\theta_{j+1}-\theta_{j}\right)}\rangle_j\rangle_\mathcal{P}|.
\end{equation} 
$\langle . \rangle_j$ denotes the average over the droplet array. The ratio $A_\Phi/A_\text{M}$ thus measures essentially the mean difference of phases between two neighboring droplets in the array. 
We also read from Eq.\,(\ref{eq:phasor}) that the phase of the phasor is
$\Phi \approx \langle \theta_{j+1}-\theta_j \rangle_j$. 

The circular variance $\Delta\Phi$ for $q$ realizations can be expressed as
\begin{equation}\label{eq_circvar_theo} 
 \Delta \Phi=1-\frac{1}{q}\sqrt{\sum_{i_1=1}^q \text{e}^{\text{i} \Phi_{i_1}} \sum_{i_2=1}^q \text{e}^{-\text{i} \Phi_{i_2}}}
\end{equation}
For a totally phase coherent state, $\Phi = 0$ for all realizations  leading to $\Delta\Phi=0$ while for a totally 
phase incoherent sample only the diagonal terms in (\ref{eq_circvar_theo}) survive, leading to $\Delta\Phi =1-\frac{1}{\sqrt{q}}$ for $q$ independent measurements.

\subsection{Modelization by a Josephson junction array}
Let us modelize the system by a set of independent droplets, each one having a number of particles $N_j$ (whose average will be denoted $\overline{N_j}$) and a phase $\theta_j$. 
The Hamiltonian of such a system is 
\begin{equation} 
H=\sum_j \left[\frac{\left(N_j-\overline{N_j}\right)^2}{2C_j}-J_j \cos\left(\theta_{j+1} - \theta_j\right)\right]
\end{equation}
The first term is the ``charging'' energy of the droplet (corresponding to its mean interaction energy) with the ``capacitance'' $C_j$ while the second term describes the Josephson tunnelling of particles between droplets with the Josephson amplitude $J_j$. This is the well known JJA description~\cite{fazio:2001}. 
Such a description is well adapted if the droplets are reasonably well separated in space, and thus should work adequately when the SSP is established. For simplicity we assume in the subsequent calculations that all parameters $C_j$ and $J_j$ are independent of the droplet and their values are later denoted $C$ and $J$ respectively. In addition, $C$ is taken as constant. 

\subsection{Time evolution for the JJA model}
To follow the experimental protocol, we used the equilibrium state of our Hamiltonian as initial state and looked at its evolution when we applied quenches to the Hamiltonian.
We took a set of $4$ droplets with, for simplicity, periodic boundary conditions.
First, we quenched our initial state with a Hamiltonian having a $J_\text{S}$ much smaller than the original $J$ for a given time $t_\text{S}$. This corresponds to a decrease of the tunneling between droplets and therefore lets them evolve independently from each other.
Then, we quenched our state again by letting it evolve a time $t_\text{h}$ with the original Hamiltonian ($J$). This means that we reinstate the original tunneling between the droplets which is what happens in the experimental protocol.

We then looked at the correlation function $|\bra{\psi\left(t\right)}\langle \text{e}^{\text{i}\theta_{j+1}}\text{e}^{-\text{i} \theta_{j}}\rangle_j\ket{\psi\left(t\right)}|$ which corresponds to $A_\Phi/A_\text{M}$. 

Given the perfect coherence that exists in the JJA model described above, this correlation function would show undamped oscillations corresponding to the time periodic nature of a system with a finite number of frequencies. In order to make the plateaus apparent in the correlation function we have damped these oscillations by an artificial damping term $\text{e}^{-\omega_k t}$ for the mode with a frequency $\omega_k$. The choice of such damping rather than the usual constant exponential one, is to get rid efficiently of the high frequency oscillations without having to recourse to the coupling to a bath for example. It is clear that a more precise and microscopically correct way of including the damping should be considered, but as discussed in the main text, what mechanism leads to damping is a whole question in itself in this system.

From time $0$ to $t_\text{s}$, this correlation function shows the decrease in $A_\Phi/A_\text{M}$ and corresponds qualitatively to the phase scrambling as can be seen in Extended Data Figure~\ref{FigS2}a.
The first minimum in the figure and the corresponding time scale would correspond to the dephasing discussed in the main text. In total absence of residual coupling between the droplets one would have lost completely the phase coherence on this timescale. Because we have put a small but finite coupling $J_\text{S}$ remaining between the droplets one can also see at later times a partial recovery of the phase coherence whose value is of course controlled by the value of $J_\text{S}$.

Furthermore, if looked up to time $t_\text{h}$, this correlation function shows the increase in $A_\Phi/A_\text{M}$ and therefore corresponds to the rephasing of the system as shown in Extended Data Figure~\ref{FigS2}b. In the calculation, the correlation does not go back to a value of $1$, which means that the rephasing is not perfect, in contrast with the experiment. This happens since this simplified model has no energy dissipation mechanism. The initial state not being an eigenstate of the final Hamiltonian thus leads to a final state which is a thermal-like state (with possibly more complicated distributions than a simple thermal one), where the extra energy has been converted to a distribution over the eigenstates. An energy dissipation mechanism, for example via the normal part of the fluid, will thus be necessary to converge back to the initial state. Such extra effects can be potentially incorporated in subsequent studies.

\begin{extfig}[t]
\centering
\includegraphics[width=\columnwidth]{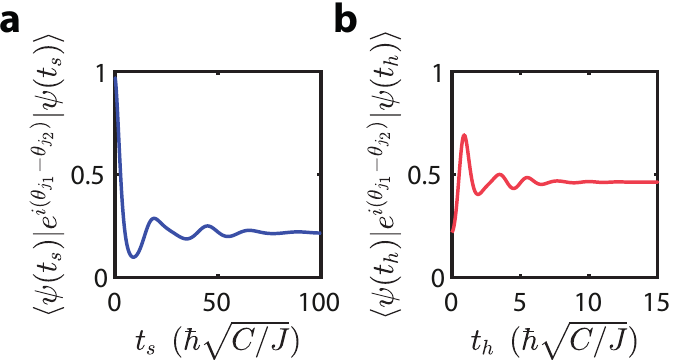} 
\caption{\label{FigS2}
\textbf{Theoretical predictions for the JJA model} with parameters $J=100$, $J_\text{S}=1$, $C=1$, $\hbar=1$ and $j_1$, $j_2$ are neighbours. \textbf{a,} Evolution of the correlation when the droplets evolve independently from each other as a function of scrambling time $t_\text{S}$. \textbf{b,} Evolution of the correlation when the droplets are re-coupled to each other as a function of $t_\text{h}$ with $t_\text{S}=1000$ $[\hbar\sqrt{\nicefrac{C}{J}}]$.}
\end{extfig}

\bibliographystylemethods{naturemag}
\bibliographymethods{references}

\end{document}